ORIGINAL RESEARCH ARTICLE

# Theory of Intermediate Twinning and Spontaneous Polarization in Ferroelectric Potassium Sodium Niobate

Georgios Grekas[1,2,3] · Patricia-Lia Pop-Ghe[1] · Eckhard Quandt[4] · Richard D. James[1]

**Abstract** Potassium sodium niobate is considered a prominent material system as a substitute for lead-containing ferroelectric materials. It exhibits first-order phase transformations and ferroelectricity with potential applications ranging from energy conversion to innovative cooling technologies, thereby addressing important societal challenges. However, a major obstacle in the application of potassium sodium niobate is its multi-scale heterogeneity and the lack of understanding of its phase transition pathway and microstructure. This can be seen from the findings of Pop-Ghe et al. (Ceram Int 47(14):20579–20585, 2021, https://doi.org/10.1016/j.ceramint.2021.04.067) which also reveal the occurrence of a phenomenon they term intermediate twinning during the phase transition. Here, we show that intermediate twinning is a consequence of energy minimization. We develop a geometrically nonlinear electroelastic energy function for potassium sodium niobate, including the cubic-tetragonal-orthorhombic transformations and ferroelectricity. The construction of the minimizers is based on compatibility conditions which ensure continuous deformations and pole-free interfaces. These minimizers agree with the experimental observations, including laminates between tetragonal variants under the cubic to tetragonal transformation, crossing twins under the tetragonal to orthorhombic transformation, intermediate twinning and spontaneous polarization. This shows how the full nonlinear electroelastic model provides a powerful tool in understanding, exploring, and tailoring the electromechanical properties of complex ferroelectric ceramics.

This invited article is part of a special issue of *Shape Memory and Superelasticity* honoring Associate Editor Petr Šittner for his 65th birthday. The special issue was organized by Hanuš Seiner, Institute of Thermomechanics, Czech Academy of Sciences.

**Supplementary Information** The online version contains supplementary material available at https://doi.org/10.1007/s40830-025-00526-z.

✉ Georgios Grekas
grekas.g@gmail.com

1 Aerospace Engineering and Mechanics, University of Minnesota, Minneapolis, USA

2 Institute of Applied and Computational Mathematics, Foundation for Research and Technology-Hellas, Heraklion, Greece

3 Computer, Electrical, Mathematical Sciences & Engineering Division, King Abdullah University of Science and Technology (KAUST), Thuwal, Saudi Arabia

4 Inorganic Functional Materials, Kiel University, Kiel, Germany

## Introduction

Ferroelectric crystals have numerous existing and potential applications, such as piezoelectric actuators, solid-state cooling devices [1–3], and devices for energy storage and conversion [4–6]. A material that exhibits superior piezoelectric and electromechanical properties is Pb(Zr,Ti)$O_3$ (PZT). Besides the exceptional ferroelectric behavior, this lead-containing state-of-the-art ferroelectric material entails critical environmental issues. As an alternative, the $K_{0.5}Na_{0.5}NbO_3$ (KNN) ceramic [7] is of increasing interest. In view of Saito's study revealing a high piezoelectric coefficient on textured KNN [8], this lead-free material is considered one of the most promising candidates in line to replace PZT [9, 10].

KNN is a complex ferroelectric ceramic: it crystallizes in a perovskite structure like barium titanate or lead zirconate titanate, but exhibits multi-scale heterogeneities [11, 12], unlike the preceding examples [13]. These multi-scale heterogeneities include a shared A-site correlated to abnormal grain growth in this compound [14], as well as different diffusion velocities and vapor pressures [15, 16] and poor sinterability [17]. Abnormal grain growth is of special importance in this context, since it suppresses the potential to accurately apply design strategies for material optimization due to the induced anisotropies. Research reported in [18] shows that an excess of alkali metals incorporated during the fabrication process suppresses this inhomogeneous grain size distribution. The resulting fatigue-optimized KNN system, called $KNN_{ex}$, exhibits two-phase transitions: orthorhombic to tetragonal at ~ 210 °C and tetragonal to cubic phases at ~ 400 °C. The latter transition is also a ferroelectric to paraelectric transition. Adding to the complexity of the material, an unexplained *intermediate twinning* state has been observed in $KNN_{ex}$ occurring during the orthorhombic to tetragonal phase transformation. The terminology refers to the bands of tetragonal variants emerging within established orthorhombic variants on heating, Fig. 1c and d. This further challenges the need for thorough understanding of phase transition dynamics in ferroelectric ceramics.

Here, we give a theoretical understanding of phase transitions, intermediate twinning [18], and spontaneous

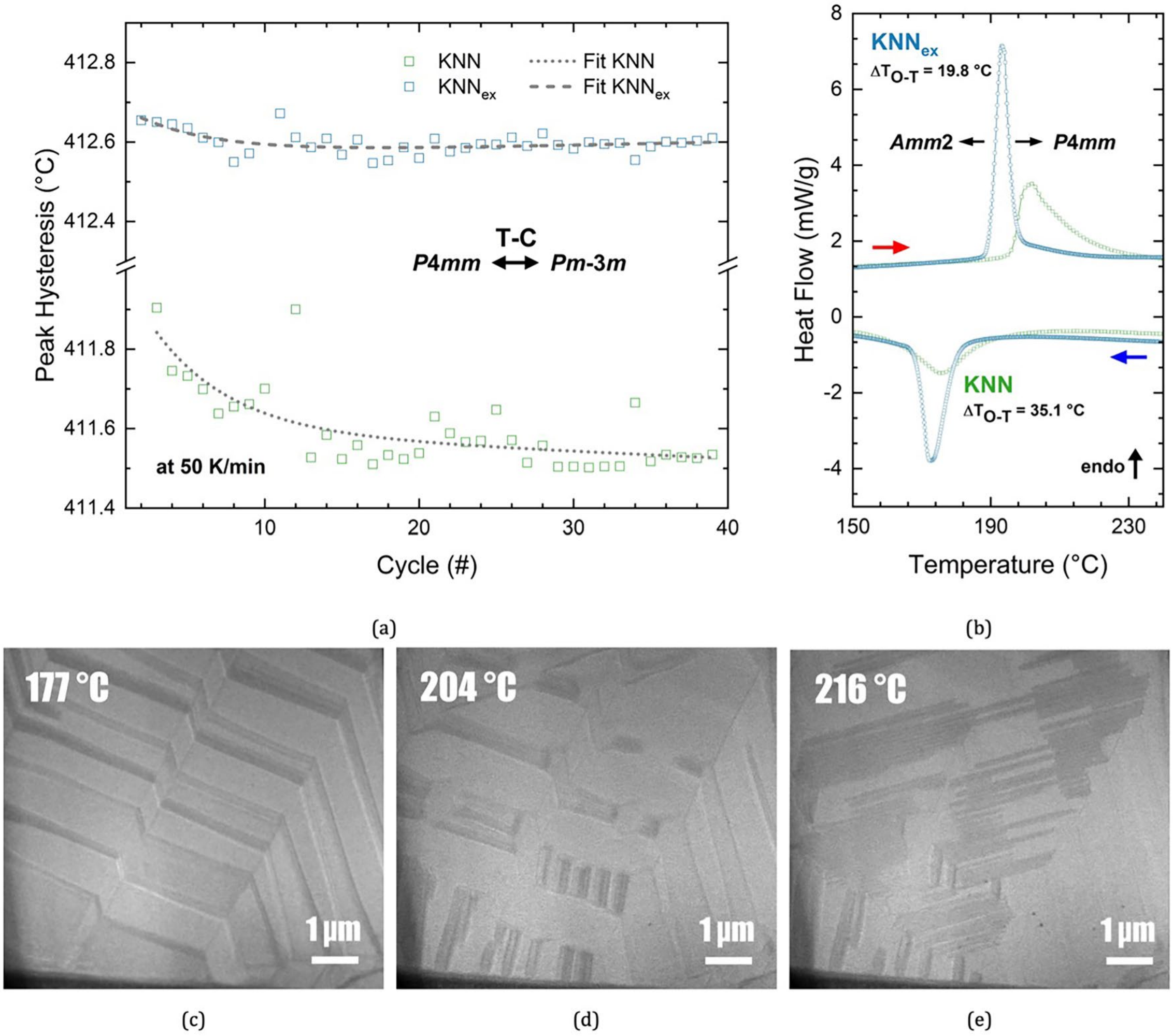

**Fig. 1** Upper panel: Differential scanning calorimetry results for **a** the tetragonal to cubic transition and **b** the orthorhombic to tetragonal transition in exemplary KNN and $KNN_{ex}$ samples, where $\lambda_{2,KNN} = 0.9974$ and $\lambda_{2,KNNex} = 0.9977$, respectively. Unit cell schematics and corresponding space groups are given in the insets, respectively. Lower panel: Intermediate twinning as discovered in [18]. Transmission electron microscopy results showing the **c** pure orthorhombic phase with orthorhombic twinning at 177 °C, **d** intermediate twinning formed within the orthorhombic variants during the O–T transition, and **e** consumption of the intermediate twins by regular tetragonal variants with completion of the phase transition; **c** and **d** are partially adapted from (Ref. [18]), available under CC BY 4.0 license at Elsevier

polarization [19, 20] in $KNN_{ex}$. We employ a nonlinear electrostrictive model derived from the related geometrically nonlinear theory of magnetostriction [21], the main difference being the possible presence of free charges in the ferroelectric case. In this theory the elastic response borrows concepts from geometrically nonlinear elasticity [22–25] and the electrical part borrows ideas from the theory of micromagnetics developed by Brown [26]. The transition from magnetostriction to electrostriction has been studied in [27] for a ferroelectric conductor system subjected to a dead load, where external mechanical and electrical work have been incorporated [21]. Examples of applications of this electroelastic model include studies of domain switching and electromechanical response in barium titanate [28–30]. In our case, the spontaneous polarization occurs due to thermally induced displacive phase transitions, which means we can directly replace magnetization by polarization to the magnetoelastic energy of [21].

In this approach the transition from the atomistic scale to the macroscopic scale is achieved via appropriate assumptions for the deformation and spontaneous polarization. The Cauchy–Born hypothesis is adopted, which links atomistic and continuum-level deformations, i.e., the lattice vectors deform as the assigned local macroscopic deformation gradient. Its validity is ensured by restricting the deformation gradient to the Ericksen–Pitteri neighborhood [23, 31, 32] which excludes plastic deformation but includes elastic deformation phase transformation and variant rearrangement. Furthermore, it is assumed the electric dipole field oscillates in a much larger scale than the scale of the lattice [33]. Then, the macroscopic polarization is derived as a volume average of the dipole field [34]. Therefore, the theoretical analysis for magnetic [21, 35] and electric [27] domain formation is inherited in the current setting.

The result is a free energy for $KNN_{ex}$ depending on $(\mathbf{F},\mathbf{p},\theta)$, where $\mathbf{F}=\nabla\mathbf{y}$ and $\mathbf{p}$ denote deformation gradient and spontaneous polarization, respectively, and $\theta$ denotes the temperature. The variants of the three phases are modeled by symmetry-related energy wells in $(\mathbf{F},\mathbf{p})$ space, which exhibit an exchange of stability at the measured transformation temperatures. X-ray diffraction measurements [18] give explicit values of the energy wells (to high accuracy) corresponding to the three phases cubic, tetragonal and orthorhombic. Energy-minimizing microstructures are found by assigning the temperature and placing the deformation gradient and corresponding polarization on the minimizing energy wells at that temperature, while ensuring that the deformation gradient is compatible and the polarization is divergence free (in the deformed configuration). Both compatibility and the divergence-free condition play an important role in the elucidation of intermediate twinning.

We show that experimentally observed domain structures, including those exhibiting intermediate twinning, minimize the total energy. In particular, our results give an explanation for the origin of intermediate twinning. Briefly, this is owing to the presence of the two transformations, together with the assumption that incipient microstructures must grow compatibly inside microstructures that are already present due to earlier transformations. A striking feature of the theory is the prediction of the interfaces between variants in the highly geometrically restrictive orthorhombic phase, together with 180°-domain formation in the same phase, Figs. 5 and 8.

In section "Experiment and Model," we present some background on the microstructure and behavior of $KNN_{ex}$, including the observed cubic, tetragonal, and orthorhombic phases and intermediate twining. Furthermore, we give some preliminary results and the definition of the electroelastic energy. In section "Results and Discussion," we present predictions of the model and comparisons with observations and experimental data.

## Experiment and Model

### Fatigue-Improved KNN

The $KNN_{ex}$ bulk samples used here exhibit a homogeneous grain size distribution [18] and optimized fatigue behavior. Preparation was executed according to the conventional solid-state route [36, 37], incorporating excess alkali metals and an optimized fabrication route [38]. The differential scanning calorimetry measurements presented in Fig. 1 show reproducible material property improvement through engineered reversibility. Figure 1a demonstrates the predicted stabilization of the phase transition temperature upon repeated cycling at accelerated degeneration conditions (cf. "Experimental Method" section) for high-purity KNN and fatigue-optimized $KNN_{ex}$ transitioning from tetragonal to cubic (T–C). The data show a qualitative difference of temperature variation, where the fatigue-optimized $KNN_{ex}$ exhibits a significantly tuned phase transition temperature as indicated by the individual two-phase exponential fitting functions. In detail, the optimized samples show a standard deviation of $\sigma=0.02872$ °C of the T–C transition temperature, while the ordinarily fabricated samples exhibit $\sigma=0.10359$ °C, translating to 0.007% and 0.025% deviation with regard to the median transition temperatures, respectively. Specifically, the first cycles are affected by a training-like effect in KNN, which is essentially neglectable for $KNN_{ex}$. Owing to a slower heating rate of 10 K $min^{-1}$, phase transition temperatures are shifted for the 12th cycle and the 11th cycle for both samples. These shifts are caused by the application of a slower heating rate in the preceding cycle [39]. Taking this into consideration, the phase transition temperature stability is more pronounced in the fatigue-optimized sample, resulting in standard deviations of 0.06% ($KNN_{ex}$) and 0.31% (KNN). As for the intermediate twinning, Fig. 1b provides details on an exemplary orthorhombic to tetragonal (O–T) transition in $KNN_{ex}$, verifying a strong improvement in $KNN_{ex}$ in terms of sharpness of the transition and amount of released latent heat, which is given by the area under the heat flow signal during the phase transition.

From an application perspective, one of the most important improvements is given by the 44% decrease of hysteresis width from $\Delta T_{KNN} = 35.1$ °C to $\Delta T_{KNNex} = 19.8$ °C, with hysteresis width being the main source of loss. Notably, this property improvement is in general more apparent in the O–T transition, exhibiting larger decrease in hysteresis width and significant sharpening of peak shape as compared to 14% decrease in the T–C transition [18]. This is interesting but indirect evidence that the O–T transition and its intermediate twinning state are important. Corresponding in situ transmission electron microscopy results of the O–T transition in $KNN_{ex}$ demonstrated that the intermediate twinning emerges with the beginning of the phase transition and vanishes completely following the tetragonal to cubic transition. The directions for spontaneous polarization in the orthorhombic KNN configuration are the ⟨110⟩ directions [19], while in the tetragonal phase, the elongation of the unit cell conditions preferred ⟨100⟩ directions for the spontaneous polarization direction [20].

## Experimental Method

The $K_{0.5}Na_{0.5}NbO_3$ (KNN) and KNN plus excess alkali metals ($KNN_{ex}$) bulk samples providing the basis for the applied model have been fabricated by solid-state route, where the detailed procedure is available elsewhere [38]. In specific, $KNN_{ex}$ samples exhibit homogeneous grain growth and have thus been chosen for comparison of the experimental results with the developed model. $KNN_{ex}$ describes samples fabricated along improved process parameters yielding suppressed abnormal grain growth and fatigue optimization with the criteria of compatibility, since the suppression of multi-scale heterogeneities allows for the accurate application of a model. Herein, $KNN_{ex}$ describes $K_{0.5}Na_{0.5}NbO_3$ samples with an alkali metal (A-site) excess of 5-mol-% potassium and 15-mol-% sodium in comparison to conventional equimolar KNN samples without any intentional deviation on the A-site. The first set of data was measured with a NETZSCH DSC 204 F1 Phoenix differential scanning calorimeter (DSC), the second set was measured with a TA Instruments Q1000. Samples were cycled at a standard rate of 10 K $min^1$ for the first and second cycle, as well as at elevated degeneration conditions of 50 K $min^1$ for the following cycles. The tangent method was used for the determination of the thermal hysteresis $\Delta T$ using the equation: $\Delta T = 1/2((A_s + A_f) - (M_s + M_f))$. Herein, $A_s$, $A_f$, $M_s$, and $M_f$ represent the respective austenitic (high-symmetry) and martensitic (low-symmetry) start and finish temperatures. With regard to reproducibility the stability of the experimental DSC parameters is verified in a preceding experiment, particularly for the heating cycle. The detailed transmission electron microscopy (TEM) results are part of a different experiment [18] and serve as an experimental verification to the developed model.

## The Electroelastic Energy

We employ the nonlinear theory of electrostriction [27] where in our setting the adopted model is derived directly from the nonlinear theory of magnetostriction [35, 40], by replacing magnetization with polarization. The total energy of the electrostatic configuration is given by

$$E(\boldsymbol{y}, \boldsymbol{p}, \theta) = \int_{\Omega} W(\nabla \boldsymbol{y}(\boldsymbol{x}), \boldsymbol{p}(\boldsymbol{y}(\boldsymbol{x})), \theta) \mathrm{d}\boldsymbol{x} + \frac{1}{2} \int_{R^3} \left| \nabla_y \phi(\boldsymbol{y}) \right|^2 \mathrm{d}\boldsymbol{y}, \tag{1}$$

where $\mathbf{y}$: $\Omega \to R^3$ is the deformation, $\mathbf{p}$: $\mathbf{y}(\Omega) \to R^3$ is the polarization, $\theta$ denotes the temperature, $W$ is the macroscopic free energy per unit volume, and $\varphi$: $\mathbf{y}(\Omega) \to R$ is the electric potential obtained by the unique solution [41], of the Maxwell's equation

$$\nabla_{\mathbf{y}} \cdot \left( \nabla_{\mathbf{y}} \varphi(\mathbf{y}) + \mathbf{p}(\mathbf{y}) \right) = 0, \quad \text{with } \varphi(\mathbf{z}) \to 0 \text{ as } |\mathbf{z}| \to +\infty. \tag{2}$$

The analog of exchange energy $\nabla \mathbf{p} \cdot \mathbf{A}\mathbf{p}$ with $\mathbf{A} \in \mathrm{R}^{3\times3}$ has been neglected in Eq. (1), under the assumption that domains are much larger than the transition layer between two distinct polarization states [42], giving rise to the possibility of piecewise uniform minimizing sequences [21, 41]. Here, the anisotropy energy $W$ models the tendency of the deformations and polarization at the preferred states and $\varphi$ is the depolarization field due to the polarization distribution in the material.

In this setting the link between the atomistic and the macroscopic deformations is provided by the Cauchy–Born hypothesis where the lattice vectors exhibit locally the same deformation as a macroscopic homogeneous deformation. The validity of the Cauchy–Born hypothesis is ensured by restricting the deformations over a neighborhood of the lattice vectors the Ericksen–Pitteri neighborhood $\mathcal{N}$, see [23, 31, 32]. Specifically, let the parent lattice $\{\mathbf{e}_i\} \in \mathcal{N}$, then the produced lattice $\{\mathbf{F}\mathbf{e}_i\}$ should also belong to the Ericksen–Pitteri neighborhood $\mathcal{N}$, where $\mathbf{F}$ is a homogeneous deformation gradient. Additional properties of $\mathcal{N}$ allow elastic deformations and phase transitions but exclude plastic deformation, i.e., dislocation motion at atomic level.

Passing to the continuum scale, the anisotropic macroscopic free energy per unit volume $W$ is obtained, where the principles of frame indifference and material symmetry are inherited by the atomistic description of the free energy

$$\begin{array}{ll} W(\mathbf{F},\mathbf{p},\theta) = W(\mathbf{RF},\mathbf{Rp},\theta), & \text{for all } \mathbf{R} \in SO(3) \text{ (Frame indifference)}, \\ W(\mathbf{F},\mathbf{p},\theta) = W(\mathbf{FQ},\mathbf{p},\theta), & \text{for all } \mathbf{Q} \in \left(\mathbf{e}_i\right) \text{ (Material Symmetry)}, \end{array} \tag{3}$$

where $\mathcal{P}(\mathbf{e}_i)$ denotes the Laue group of the lattice $\{\mathbf{e}_i\}$. Furthermore, it is assumed that there exists a critical temperature $\theta_c$ such that the free energy density $W$ is minimized for $(\boldsymbol{F},\boldsymbol{p}) = \left(1,\boldsymbol{p}_1\right)$, the parent phase, when $\theta \geq \theta_c$ and is minimized for $(\boldsymbol{F},\boldsymbol{p}) = \left(\boldsymbol{U},\boldsymbol{p}_U\right)$, the daughter phase, when $\theta \leq \theta_c$. It should be noted if $\left(\boldsymbol{U}_1,\boldsymbol{p}_1\right)$ is a minimizer of $W$ then due to frame indifference and material symmetry $\left(\mathbf{R}\boldsymbol{U}_1 Q, \mathbf{R}\boldsymbol{p}_1\right)$ is also a minimizer, for all $\mathbf{R} \in SO(3)$ and $\mathbf{Q} \in \mathcal{P}\left(U_1 e_i\right)$, where $\{\mathbf{e}_i\}$ denotes the parent lattice. The set containing all the minimizers ("energy wells") at temperature $\theta$ is defined as

$$\mathcal{M}_\theta = \cup_{i=1}^{n} SO(3)\{\boldsymbol{U}_i(\theta), \pm\boldsymbol{p}_i(\theta)\}, \tag{4}$$

where $\boldsymbol{U}_i(\theta) = \boldsymbol{Q}_i^T \boldsymbol{U}_1(\theta)\boldsymbol{Q}_i$, for $i = 1,..,n$ and $\mathbf{Q} \in \mathcal{P}\left(\boldsymbol{U}_1(\theta)\boldsymbol{e}_i\right)$, denote the $n$ distinct variants of the phase at temperature $\theta$. Without loss of generality, we assume $W(\boldsymbol{A},\boldsymbol{p}_A,\theta) = 0$ if $(\boldsymbol{A},\boldsymbol{p}_A) \in \mathcal{M}_\theta$. The energy wells of $\mathcal{M}_\theta$ can be restricted to satisfy the saturation hypothesis $|\det\mathbf{F}(\mathbf{x})\mathbf{p}(\mathbf{y}(\mathbf{x}))| = g(\theta)$ [21], for some $g$: $R \rightarrow R^+$.

The preceding paragraph applies to the case of a single transformation. In the present case, there are two such temperatures, $\theta_c^{c\rightarrow t} > \theta_c^{t\rightarrow o}$, corresponding to the two transformations, and the Ericksen–Pitteri neighborhoods are nested, with the larger one corresponding to the cubic lattice. The considerations above apply to each of these transformations.

## Minimization of the Total Energy

We study minimizers of $E(\boldsymbol{y},\boldsymbol{p},\theta)$ given by (1) at various temperatures, with special attention to the two temperatures $\theta_c^{c\rightarrow t}$ and $\theta_c^{t\rightarrow o}$. Since the general considerations of this subsection apply in both cases, we simply assume the energy-well configuration defined by (4).

Coexistence of the phases has been observed in $\text{KNN}_{\text{ex}}$ [18]. The deformations that minimize the nonlinear elastic energy $W$ should be invertible and continuous but may have a discontinuous deformation gradient, and the values of the deformation gradient must belong to the bottoms of the energy wells, i.e., $\mathcal{M}_\theta$ given by (4). To ensure the resulting deformations are continuous, they should satisfy the Hadamard jump condition. Specializing the condition, let a region be divided by a plane with normal $\mathbf{n}$. Deforming the regions, assume the deformation gradient takes the values $\mathbf{RB}$ and $\mathbf{A}$ on opposite sides of a planar interface, for some $\mathbf{R} \in SO(3)$. Then, the deformation $\mathbf{y}$ is continuous (termed "compatible") if and only if the twinning equation

$$\mathbf{RB} - \mathbf{A} = \mathbf{a} \otimes \mathbf{n} \tag{5}$$

holds for some $\mathbf{a} \in R^3$ and $|\mathbf{n}| = 1$. Here, $\boldsymbol{a} \otimes \boldsymbol{n}$ is a $3 \times 3$ matrix with components $(\boldsymbol{a} \otimes \boldsymbol{n})_{ij} = \boldsymbol{a}_i\boldsymbol{n}_j$, i.e., $\mathbf{A}$ and $\mathbf{RB}$ must be "rank-one connected." Ball and James [22] provide necessary and sufficient conditions for the solution of (5). Specifically, given $\mathbf{A}$ and $\mathbf{B}$, they proved that if the middle eigenvalue of the matrix $C = F^{\mathrm{T}}F$ is one, where $F = BA^{-1}$, then there exist two solutions of (5) given in [22] and denoted by $\mathbf{R}^{\pm}$,$\mathbf{a}^{\pm}$,$\mathbf{n}^{\pm}$, with $\mathbf{R}^{\pm}$ in $SO(3)$. The two solutions corresponding to $\pm$ are called reciprocal twins.

For the same region let $\boldsymbol{p}_A$, $\boldsymbol{p}_B$ denote the polarization vectors in the deformed configuration of the phases corresponding to $\mathbf{A}$, $\mathbf{B}$, respectively. The depolarization energy is minimized when $\nabla\varphi = \mathbf{0}$, but from Eq. (2) this occurs when $\nabla \cdot \mathbf{p} = 0$ for the interior points of the deformed body and when the polarization vector is perpendicular to the normal of $\partial y(\Omega)$. Assuming that $\nabla \cdot \boldsymbol{p}_A = 0$, $\nabla \cdot \boldsymbol{p}_B = 0$, the extra compatibility condition is

$$\left(\boldsymbol{p}_A - \boldsymbol{p}_B\right) \cdot \boldsymbol{m} = 0, \text{where} \boldsymbol{m} = \frac{\boldsymbol{A}^{-\mathrm{T}}\boldsymbol{n}}{\left|\boldsymbol{A}^{-\mathrm{T}}\boldsymbol{n}\right|} = \frac{\boldsymbol{B}^{-\mathrm{T}}\boldsymbol{n}}{\left|\boldsymbol{B}^{-\mathrm{T}}\boldsymbol{n}\right|}, \tag{6}$$

i.e., $\mathbf{m}$ is the unit normal of the planar interface in the deformed configuration. Equation (6) ensures a divergence-free polarization at this plane implying a pole-free interface, see [21, 27]. When interfaces are formed between different variants of the same phase, these conditions are simplified due to the following lemma:

**Lemma 1** (Lemma 6.1 from [21]) *Suppose* $\mathbf{A}$ *and* $\mathbf{B}$ *are symmetric matrices with* $\det\mathbf{A} = \det\mathbf{B}$ *such that*

$$\mathbf{RB} - \mathbf{A} = \mathbf{a} \otimes \mathbf{n}, \text{ for some. } \mathbf{R} \in SO(3), \mathbf{a}, \mathbf{n} \in \mathrm{R}^3. \tag{7}$$

*If* $\mathbf{A}\boldsymbol{p}_A = \alpha\boldsymbol{p}_A$ *and* $\mathbf{B}\boldsymbol{p}_B = \alpha\boldsymbol{p}_B$ *for some* $\alpha \in R$, *then*

$$\left(\mathbf{R}\boldsymbol{p}_B - \boldsymbol{p}_A\right) \cdot \mathbf{m} = 0 \Leftrightarrow \left(\boldsymbol{p}_B - \boldsymbol{p}_A\right) \cdot \mathbf{n} = 0, \tag{8}$$

$\boldsymbol{m} = \boldsymbol{A}^{-1}\boldsymbol{n}$ *is the normal in the deformed configuration.*

Based on the compatibility conditions (5) and (8) appropriate deformations and polarization vectors $(\mathbf{y},\mathbf{p})$ minimizing the total energy will be constructed. There are some cases in which there is no minimizer, but there is a lower bound to the energy together with sequences $\left(\boldsymbol{y}_k,\boldsymbol{p}_k\right)$, $k = 1,2,\ldots$, whose energy converges of this lower bound

**Table 1** Tetragonal variants under cubic to tetragonal transformations

$$\boldsymbol{U}_1 = \begin{pmatrix} \gamma_t & 0 & 0 \\ 0 & \alpha_t & 0 \\ 0 & 0 & \alpha_t \end{pmatrix}, \quad \boldsymbol{U}_2 = \begin{pmatrix} \alpha_t & 0 & 0 \\ 0 & \gamma_t & 0 \\ 0 & 0 & \alpha_t \end{pmatrix}, \quad \boldsymbol{U}_3 = \begin{pmatrix} \alpha_t & 0 & 0 \\ 0 & \alpha_t & 0 \\ 0 & 0 & \gamma_t \end{pmatrix}$$

as $k \to \infty$, i.e., minimizing sequences. In a case without polarization an example of the latter is the classic austenite/twinned martensite interface. Below, we find some cases of this type.

## Results and Discussion

### The Cubic-to-Tetragonal Transformation

We first show that microstructures involving the phase transformation at $\theta^{c\to t}$ in $KNN_{ex}$ can be modeled via minimizing sequences of the proposed macroscopic electroelastic energy (1). We ignore polarization in the first part for simplicity; later, we insert polarization vectors and verify using Lemma 1 that these produce pole-free microstructures. In the following, $\{\boldsymbol{e}_1, \boldsymbol{e}_2, \boldsymbol{e}_3\}$ denote orthonormal vectors defining the unit cell for the cubic phase and the components of all matrices and vectors will be written in this basis. The tetragonal phase of $KNN_{ex}$ has stretch tensors of the classic form written in this cubic basis, Table 1. These three matrices correspond to the three variants of martensite. The measured values from Table S1 are used below: $\alpha_t = A_t/A_c = 3.96/3.98$ and $\gamma_t = C_t/A_c = 4.02/3.98$. Note that $\alpha_t \neq 1$ which means $U_i$ is not rank-one connected to **1** and (5) has no solutions. Instead, the cubic to tetragonal transformation is formed though a more complicated microstructure.

The construction of microstructures involving these three variants and austenite is a classic problem in the Crystallographic Theory of Martensite. We begin by constructing simple laminates of two tetragonal variants with deformation gradients $\mathbf{U}_1$ and $\mathbf{RU}_2$. Ball and James [22, 23] presented necessary and sufficient conditions for the existence of laminates between two variants providing explicit forms. Namely, under the cubic to tetragonal transitions such laminates exists if and only if the parameters of Eq. (1) satisfy

$$\alpha_t < 1 < \gamma_t \quad \text{and} \quad \frac{1}{\alpha_t^2} + \frac{1}{\gamma_t^2} < 2. \tag{9}$$

Indeed, the above inequalities hold for the tetragonal phase of $KNN_{ex}$. Then to construct the laminate, we proceed as follows. First, the twinning equation $\mathbf{RU}_2 - \mathbf{U}_1$ is solved through [22, Proposition 4], which has the solutions $(\mathbf{R}^+, \mathbf{a}^+, \mathbf{n}^+)$ and $(\mathbf{R}^-, \mathbf{a}^-, \mathbf{n}^-)$, where

$$\begin{aligned} &\boldsymbol{n}^{\pm} = \frac{1}{\sqrt{2}}(1, \pm 1, 0), \quad \boldsymbol{a}^{\pm} = \frac{\sqrt{2}(a_t - c_t)}{a_t^2 + c_t^2}(\pm c_t, a_t, 0), \\ &\boldsymbol{R}^{\pm} = \left(\boldsymbol{U}_1 + \boldsymbol{a}^{\pm} \otimes \boldsymbol{n}^{\pm}\right)\boldsymbol{U}_2^{-1}. \end{aligned} \tag{10}$$

The macroscopic deformations $\boldsymbol{A}^{\pm}_{\lambda} = \lambda \boldsymbol{R}^{\pm}\boldsymbol{U}_2 + (1-\lambda)\boldsymbol{U}_1$ are compatible to **1**, see [22, Theorem 7], for $\lambda = \lambda^*$ or $\lambda = 1 - \lambda^*$, where

$$\lambda^* = \frac{1}{2}\left(1 - \sqrt{1 + \frac{2}{\delta^{\pm}}}\right), \quad \text{and } \delta^{\pm} = \boldsymbol{a}^{\pm} \cdot \boldsymbol{U}_1\left(\boldsymbol{U}_1 - 1\right)^{-1}\boldsymbol{n}^{\pm}. \tag{11}$$

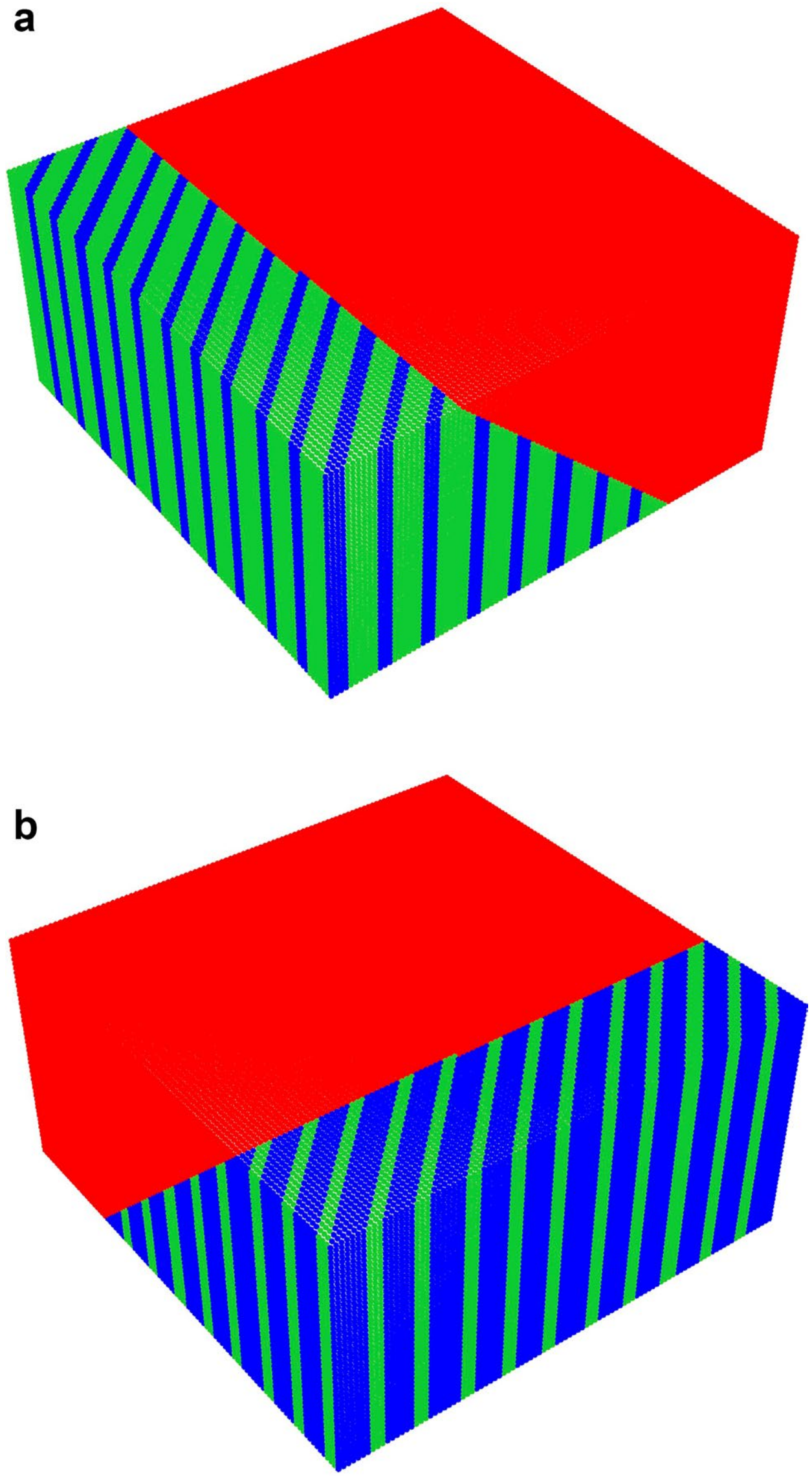

**Fig. 2** Cubic to tetragonal transformation drawn with measured lattice parameters of $KNN_{ex}$. A twinned laminate $(\boldsymbol{A}_\lambda)$ between the wells $\boldsymbol{RU}_2$ (blue color) and $\boldsymbol{U}_1$(green color) compatible to the cubic phase (red color) as the fineness parameter $k \to \infty$. **a** and **b** correspond to the laminates $\boldsymbol{A}_{\lambda^*}$ and $\boldsymbol{A}_{1-\lambda^*}$. The volume fractions of $\boldsymbol{RU}_2$ and $\boldsymbol{U}_1$ are $\lambda^*$, $1-\lambda^*$, respectively. Here, $\boldsymbol{RU}_2 - \boldsymbol{U}_1 = \boldsymbol{a} \otimes \boldsymbol{n}$ and the width of each $\boldsymbol{RU}_2$ strip along direction **n** is $\lambda^*/k$ and $(1-\lambda^*)/k$ for each $\boldsymbol{U}_1$ strip, $\lambda^* \approx 0.335047$. A transition layer of width $d/k$ for some $d > 0$ is formed, not visible here. In total, there are 24 possible interfaces of this type, see Eqs. (10), (12), and (13) (Color figure online)

Note there are four solutions resulting from the combinations

**Table 2** Orthorhombic variants under cubic-to-orthorhombic transformations

$$V_1 = \begin{pmatrix} \gamma_0 & 0 & 0 \\ 0 & \frac{\alpha_0+\beta_0}{2} & \frac{\alpha_0-\beta_0}{2} \\ 0 & \frac{\alpha_0-\beta_0}{2} & \frac{\alpha_0+\beta_0}{2} \end{pmatrix}, \quad V_2 = \begin{pmatrix} \gamma_0 & 0 & 0 \\ 0 & \frac{\alpha_0+\beta_0}{2} & \frac{\beta_0-\alpha_0}{2} \\ 0 & \frac{\beta_0-\alpha_0}{2} & \frac{\alpha_0+\beta_0}{2} \end{pmatrix},$$

$$V_3 = \begin{pmatrix} \frac{\alpha_0+\beta_0}{2} & 0 & \frac{\alpha_0-\beta_0}{2} \\ 0 & \gamma_0 & 0 \\ \frac{\alpha_0-\beta_0}{2} & 0 & \frac{\alpha_0+\beta_0}{2} \end{pmatrix}, \quad V_4 = \begin{pmatrix} \frac{\alpha_0+\beta_0}{2} & 0 & \frac{\beta_0-\alpha_0}{2} \\ 0 & \gamma_0 & 0 \\ \frac{\beta_0-\alpha_0}{2} & 0 & \frac{\alpha_0+\beta_0}{2} \end{pmatrix}$$

$$V_5 = \begin{pmatrix} \frac{\alpha_0+\beta_0}{2} & \frac{\alpha_0-\beta_0}{2} & 0 \\ \frac{\alpha_0-\beta_0}{2} & \frac{\alpha_0+\beta_0}{2} & 0 \\ 0 & 0 & \gamma_0 \end{pmatrix}, V_6 = \begin{pmatrix} \frac{\alpha_0+\beta_0}{2} & \frac{\beta_0-\alpha_0}{2} & 0 \\ \frac{\beta_0-\alpha_0}{2} & \frac{\alpha_0+\beta_0}{2} & 0 \\ 0 & 0 & \gamma_0 \end{pmatrix}$$

$$\begin{aligned} &(\boldsymbol{R}^+, \boldsymbol{a}^+, \boldsymbol{n}^+, \lambda^*), \quad (\boldsymbol{R}^+, \boldsymbol{a}^+, \boldsymbol{n}^+, 1-\lambda^*), \\ &(\boldsymbol{R}^-, \boldsymbol{a}^-, \boldsymbol{n}^-, \lambda^*), \quad (\boldsymbol{R}^-, \boldsymbol{a}^-, \boldsymbol{n}^-, 1-\lambda^*). \end{aligned} \tag{12}$$

Furthermore, we compute the twinning equation for the laminate with the cubic phase, where for every choice from the above equations, we obtain

$$\boldsymbol{Q}_i\boldsymbol{A}^{\pm}_{\lambda^*} - 1 = \boldsymbol{b}_i \otimes \boldsymbol{m}_i, \overline{\boldsymbol{Q}}_i\boldsymbol{A}^{\pm}_{1-\lambda^*} - 1 = \overline{\mathbf{b}}_i \otimes \overline{\mathbf{m}}_i, \quad i = 1 \text{ or } 2, \tag{13}$$

which means there are 8 possible laminates between $\mathbf{U}_1$ and $\mathbf{U}_2$ compatible with $\mathbf{1}$. Including also $\mathbf{U}_3$, there are overall 24 possible cubic/twinned tetragonal interfaces. For Fig. 2, we have chosen the solutions corresponding to twinned laminates $\boldsymbol{A}^+_{\lambda^*}$ and $\boldsymbol{A}^+_{1-\lambda^*}$.

Essentially, $\mathbf{A}_\lambda$ can be seen as the gradient of the macroscopic deformation, i.e., the limit of a minimizing sequence of deformations $\{\mathbf{y}_k\}$ in the martensitic region as $k \to +\infty$, or, technically, $\mathbf{y}_k \overset{*}{\rightharpoonup} \mathbf{y}$ in $W^{1,\infty}(\Omega)^3$ and $\mathbf{y}(\mathbf{x}) = \mathbf{A}_\lambda \mathbf{x}$. In this sequence, $\lambda$ denotes the volume fraction of $\mathbf{QRU}_2$ and $1-\lambda$ the volume fraction of $\mathbf{QU}_1$, $\mathbf{Q}$ is the rotation satisfying Eq. (13), while $k$ denotes the fineness of each variant. An instance of this minimizing sequence is shown in Fig. 2.

The transition layer between the laminate and the austenite phase is not shown (and would hardly be visible with these material parameters). But as $k \to +\infty$ it is standard to show that the energy of this layer tends to 0 [22–24, 43, 44]. It is well accepted that the fineness of the laminate is determined by the competition between a small interfacial energy on the twin boundaries and the elastic energy in this transition layer, but that larger-scale features, such as the volume fraction $\lambda^*$ and geometry are predicted accurately by this minimizing sequence, this being the essence of the Crystallographic Theory of Martensite [45, 46].

In summary, at Curie temperature there exist a sequence of deformations such that

$$\lim_{k\to+\infty} E[\mathbf{y}_k, 0] = \lim_{k\to+\infty} \int_\Omega \phi(\mathbf{y}_k, 0)\mathrm{d}x = 0. \tag{14}$$

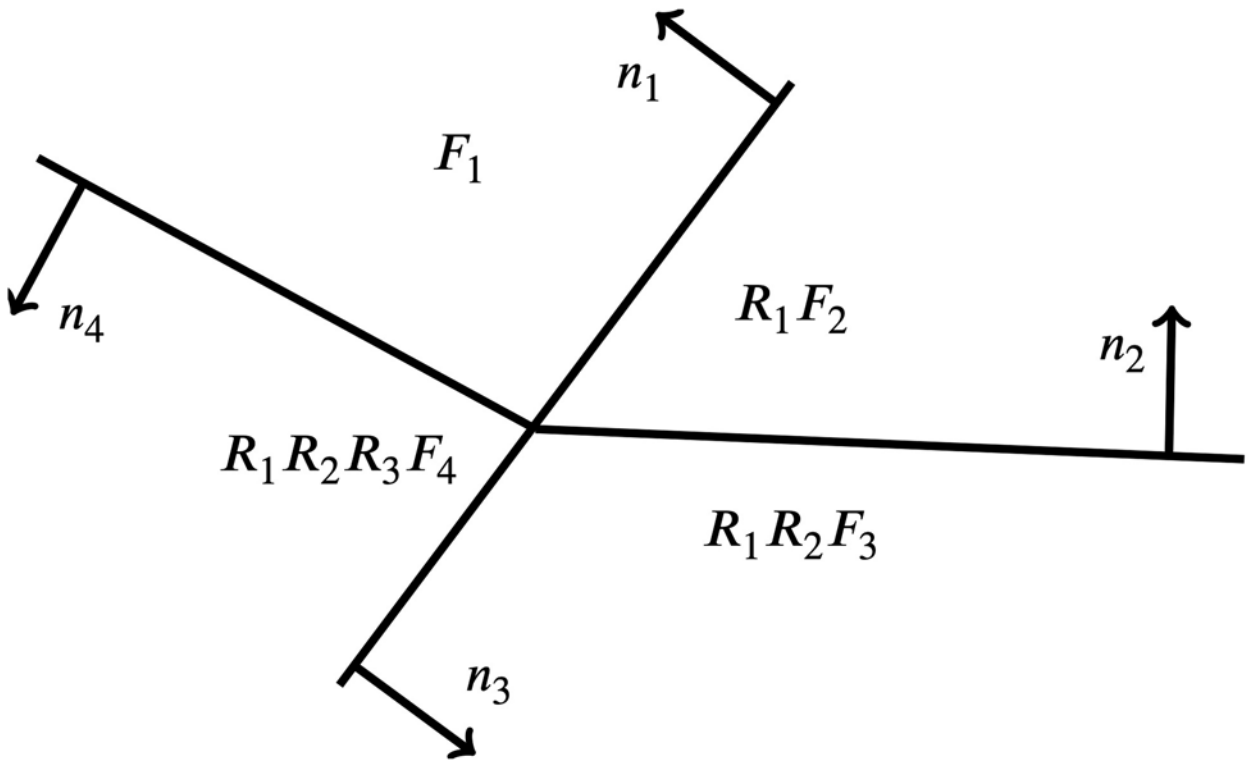

**Fig. 3** The fourfold structure formed from a crossing twin in the reference configuration

## The Orthorhombic Phase

We now consider the pure orthorhombic phase, shown experimentally in Fig. 1. As seen in this figure, more complicated microstructures appear, especially parallelogram regions that meet along a line. We assume these microstructures involve four orthorhombic variants as in some shape memory alloys, where this structure is quite common and is known as a simple crossing twin or a parallelogram microstructure [40, 43, 47–49]. Despite the fact that the kinematic constraints for the formation of crossing twins are severe, we show the microstructure is a consequence of energy minimization for the measured lattice parameters of $KNN_{ex}$ assuming the lattice correspondence

$$[100]_t \to [100]_o, [011]_t \to [010]_o, \left[0\bar{1}1\right]_t \to [001]_o.$$

The best fitting with experiments has been achieved using the above relations (see Fig. 5 and supplement), which are different than the previously reported correspondences for ceramics [50]. Therefore, the orthorhombic variants, Table 2, obtain the values $\alpha_o = \frac{A_0}{A_c}, \beta_o = \frac{B_0}{\sqrt{2}A_c}, \gamma_0 = \frac{C_o}{\sqrt{2}A_c}$, see Table S2 for the values of the measured lattice parameters $A_0, B_0, C_0, A_c$.

Let four distinct orthorhombic variants form a crossing twin (Fig. 3). From the twinning Eq. (5) two variants sharing the interface with normal $\boldsymbol{n}_i$ should be compatible. These rank-one connections do not suffice for a continuous deformation because it should be compatible also at the corners, i.e., the line where the four variants meet. Then, geometric compatibility requires the following

$$\begin{aligned} &R_1F_2 - F_1 = a_1 \otimes n_1, \quad R_2F_3 - F_2 = a_2 \otimes n_2 \\ &R_3F_4 - F_3 = a_3 \otimes n_3, \quad R_4F_1 - F_4 = a_4 \otimes n_4 \quad , \\ &R_1R_2R_3R_4 = 1, \text{ with } R_i \in SO(3), \ n_1, n_2, n_3, n_4 \text{ coplanar,} \end{aligned} \tag{15}$$

where $R_4 = \left(R_1R_2R_3\right)^{\mathrm{T}}$. In all cases here and below we choose the reference configuration $\Omega$ to be the undistorted cubic phase. The latter conditions are that the product of rotations is the identity and the normal vectors of the interfaces should be coplanar. The former requirement prevents dislocations at the corners, while the latter imposes that all four interfaces meet along a line. The majority of the martensitic transformations involve type I and type II twins [24]. These are twins between variants **A** and **B** that are related by a 180° rotation $Q_\pi(v) \in \mathcal{P}(e_i) \setminus \mathcal{P}(Ae_i)$, where **A** is a martensitic variant, $Q_\pi(v)$ denotes 180° rotation about the axis **v** and $B = Q_\pi(v)AQ_\pi(v)$. For this fourfold arrangement of twins, we have the following result from [47].

**Theorem 1** (Theorem 2 in [47]) *Let* $Q_1 = Q_\pi(v_1)$ *and* $Q_2 = Q_\pi(v_2)$ *for* $v_1 \cdot v_2 = 0$, *if*

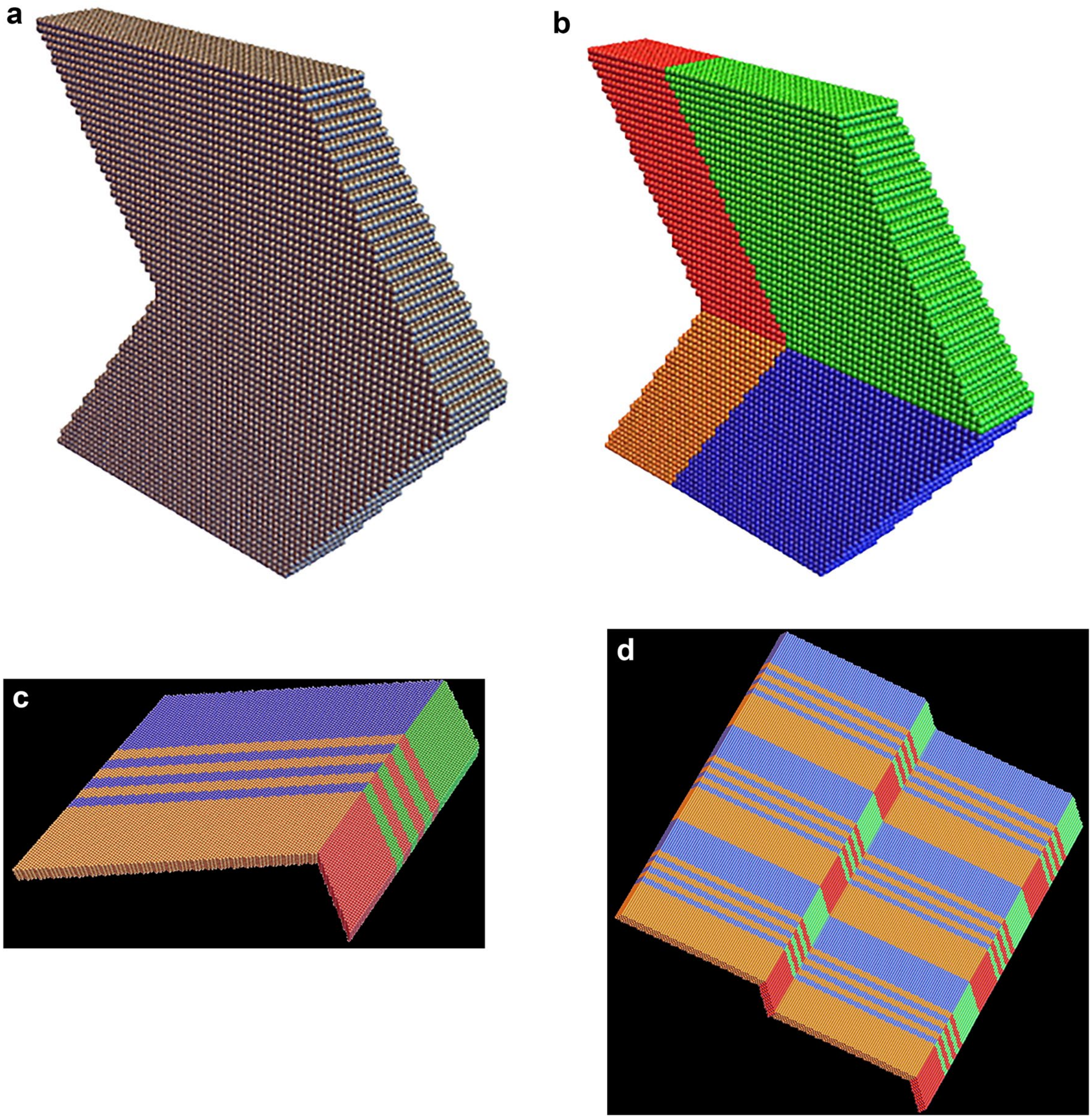

**Fig. 4** Formation of orthorhombic crossing twins. **a** Cubic reference configuration. **b** Orthorhombic deformed configuration. The variants $V_3, V_4, V_5$ and $V_6$ correspond to orange, red, green, and blue coloring, respectively. Consequently, the orthorhombic crossing twins in the C–T transition can be constructed. First, the microstructure of **c** is created and subsequently extended periodically in **d**. The variants in **d** look like they are bending far out-of-plane but actually they are nearly in-plane as in **b**. Here, in Table 2, $\alpha_o \simeq 0.9899$ and $\beta_o \simeq 1.002\gamma_o \simeq 1.007$, from XRD measurement see Table S1 (Color figure online)

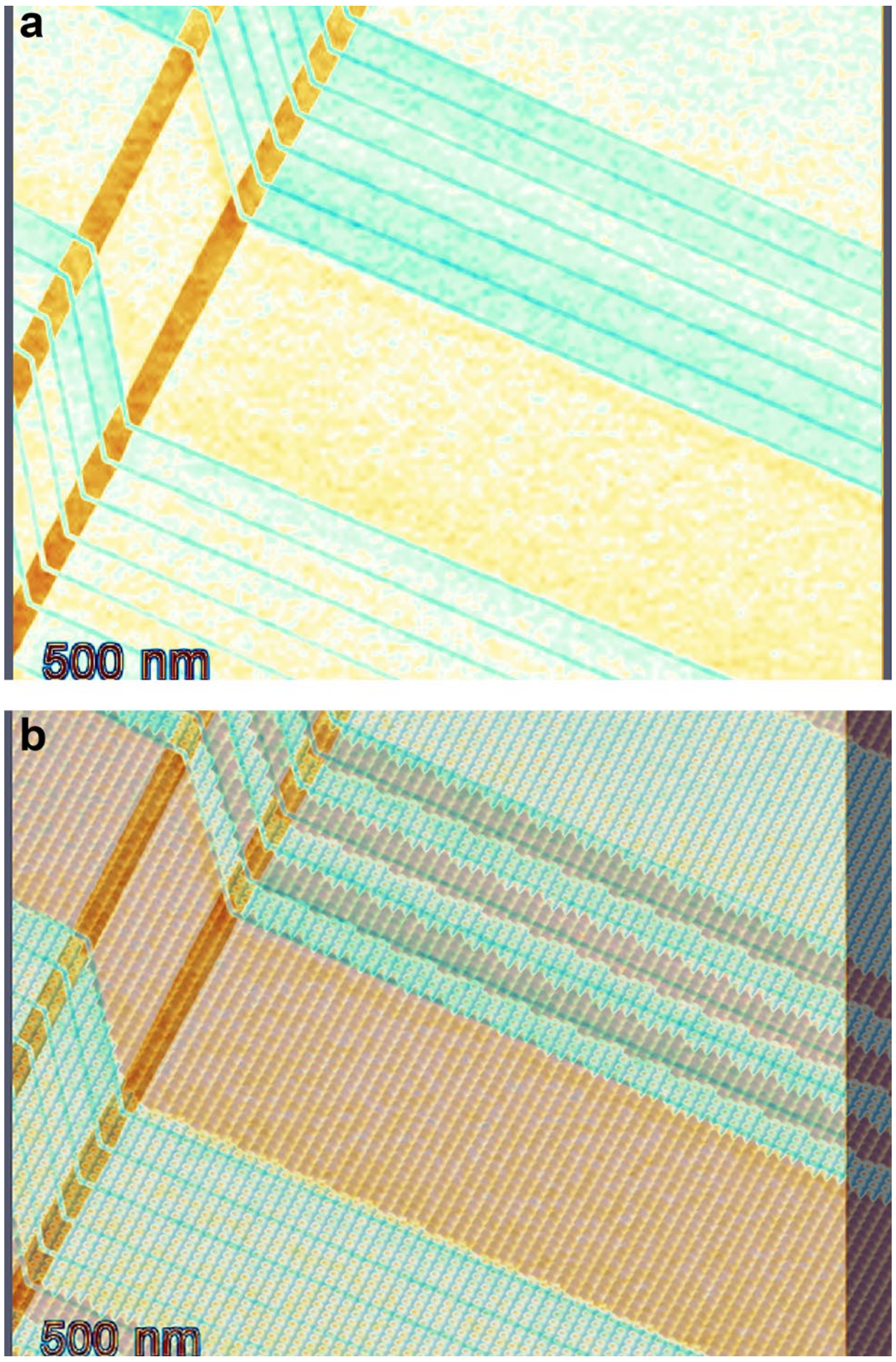

**Fig. 5** Comparing experimental observed and theoretical predicted interfaces between orthorhombic variants. **a** Artificial colored STEM image of the orthorhombic phase distinguishing different orthorhombic variants, reproduced from [18]. The different contrast originates from a different fraction of diffracted intensity for each variant. **b** Experimental image is superposed with the theoretical computed crossing twin from Fig. 4c. The theoretical interfaces coincide with the experimental observation. Figure **a** is partially adapted from (Ref. [18]), available under CC BY 4.0 license at Elsevier

$$\boldsymbol{F}_2 = \boldsymbol{Q}_1\boldsymbol{F}_1\boldsymbol{Q}_1, \boldsymbol{F}_3 = \boldsymbol{Q}_2\boldsymbol{F}_2\boldsymbol{Q}_2, \boldsymbol{F}_4 = \boldsymbol{Q}_2\boldsymbol{F}_1\boldsymbol{Q}_2, \tag{16}$$

*then the crossing twins Eq. (15) has a solution and* $R_i, a_i, n_i$ *are given explicitly.*

Assuming that Type I and Type II twins are involved, applying the above theorem we obtain the crossing twin microstructure for the orthorhombic phase of $KNN_{ex}$ as follows. Let $\mathbf{V}_i$, $i = 1,\ldots,6$ denote the orthorhombic variants as shown in Table 2. Fixing an orthorhombic variant we would like to find $180^o$ rotations from the set $\mathcal{P}(\boldsymbol{e}_i)\setminus\mathcal{P}(\boldsymbol{V}_1\boldsymbol{e}_i)$ such that Eq. (16) holds. Setting $\boldsymbol{F}_1 = \boldsymbol{V}_5$, $\boldsymbol{Q}_1 = \boldsymbol{Q}_\pi(\boldsymbol{e}_1)$, $\boldsymbol{Q}_2 = \boldsymbol{Q}_\pi(\boldsymbol{e}_2 + \boldsymbol{e}_3)$, (16) is satisfied for $\boldsymbol{F}_2 = \boldsymbol{V}_6$, $\boldsymbol{F}_3 = \boldsymbol{V}_3$ and $\boldsymbol{F}_4 = \boldsymbol{V}_4$, and $\boldsymbol{R}_i\boldsymbol{a}_i$ and $\boldsymbol{n}_i$ are computed from [47, Theorem 2]. Note that one can choose $\boldsymbol{Q}_1 = \boldsymbol{Q}_\pi(\boldsymbol{e}_2 + \boldsymbol{e}_3)$, $\boldsymbol{Q}_2 = \boldsymbol{Q}_\pi(\boldsymbol{e}_1)$ resulting in a different crossing twin. We have chosen the former rotations due to better agreement with experiments. Recalling that reference configuration $\Omega$ is chosen as the undistorted cubic phase, we define the deformation $\boldsymbol{y}(\boldsymbol{x}) = \boldsymbol{B}(\boldsymbol{x})\boldsymbol{x}, \boldsymbol{x} \in \Omega$ by

$$\boldsymbol{B}(\boldsymbol{x}) = \begin{cases} \boldsymbol{V}_5, \textit{if } \boldsymbol{x}\cdot\boldsymbol{n}_1 \geq 0 \textit{ and } \boldsymbol{x}\cdot\boldsymbol{n}_4 < 0, \\ \boldsymbol{R}_1\boldsymbol{V}_6, \textit{if } \boldsymbol{x}\cdot\boldsymbol{n}_2 \geq 0 \textit{ and } \boldsymbol{x}\cdot\boldsymbol{n}_1 < 0, \\ \boldsymbol{R}_1\boldsymbol{R}_2\boldsymbol{V}_3, \textit{if } \boldsymbol{x}\cdot\boldsymbol{n}_3 \geq 0 \textit{ and } \boldsymbol{x}\cdot\boldsymbol{n}_2 < 0, \\ \boldsymbol{R}_1\boldsymbol{R}_2\boldsymbol{R}_3\boldsymbol{V}_4, \textit{if } \boldsymbol{x}\cdot\boldsymbol{n}_4 \geq 0 \textit{ and } \boldsymbol{x}\cdot\boldsymbol{n}_3 < 0. \end{cases} \tag{17}$$

For this deformation the twins with interfaces between $\boldsymbol{V}_5$, $\boldsymbol{V}_6$ and $\boldsymbol{V}_3$, $\boldsymbol{V}_4$ are compound twins and for interfaces between $\boldsymbol{V}_4$, $\boldsymbol{V}_5$ and $\boldsymbol{V}_3$, $\boldsymbol{V}_6$ are of Type II, see also [47]. The deformation $\mathbf{y}$ is continuous, piecewise homogeneous and minimizes the free energy, i.e., $(\nabla\mathbf{y},\mathbf{0}) \in \mathcal{M}_{\theta_0}$, $\theta_0 = 187^o$ (orthorhombic phase in Fig. 1). In Fig. 4a and b, the reference state (cubic) and the deformed configuration of orthorhombic crossing twins are illustrated. More complicated microstructures are formed in Fig. 4c(S1a) by periodic extension, Fig. 4d(S1b), [51, 52]. Here, the reference configuration and the deformation are extended using a translation along the direction $\boldsymbol{n}_1$ and repeating the procedure a second extension is performed along direction $(\boldsymbol{n}_1 \times \boldsymbol{n}_2) \times \boldsymbol{n}_1$ (see subsection Extending energy-minimizing deformations in supplementary material). The extended deformation is continuous and energy minimizing.

A striking feature of this microstructure concerns the angles formed between the interfaces of the crossing twins. An image obtained from scanning transmission electron microscopy (STEM), colored to emphasize the interfaces between the orthorhombic variants, is presented in Fig. 5a. The theoretically predicted microstructure of Fig. 4d is then superposed onto this experimentally observed microstructure. The result is Fig. 5b. As can be seen, the registry of interfaces at this scale is nearly perfect. Here, the narrow bands of Fig. 5a indicate alternating orthorhombic variants. Recall that the only input for the nonlinear elastic energy are values of the energy wells obtained from the XRD measurement. Producing compatible deformations in the orthorhombic phase the remarkable agreement between experiments and theory is achieved.

## Intermediate Twinning

Intermediate twinning refers to the emergence of bands of tetragonal variants within established orthorhombic variants as the temperature is raised, Fig. 1. The opposite process—emergence of a variety of orthorhombic variants within tetragonal ones on cooling—is treated above, and our analysis shows that there is a variety of compatible interfaces. The key question is then whether these interfaces can be arranged

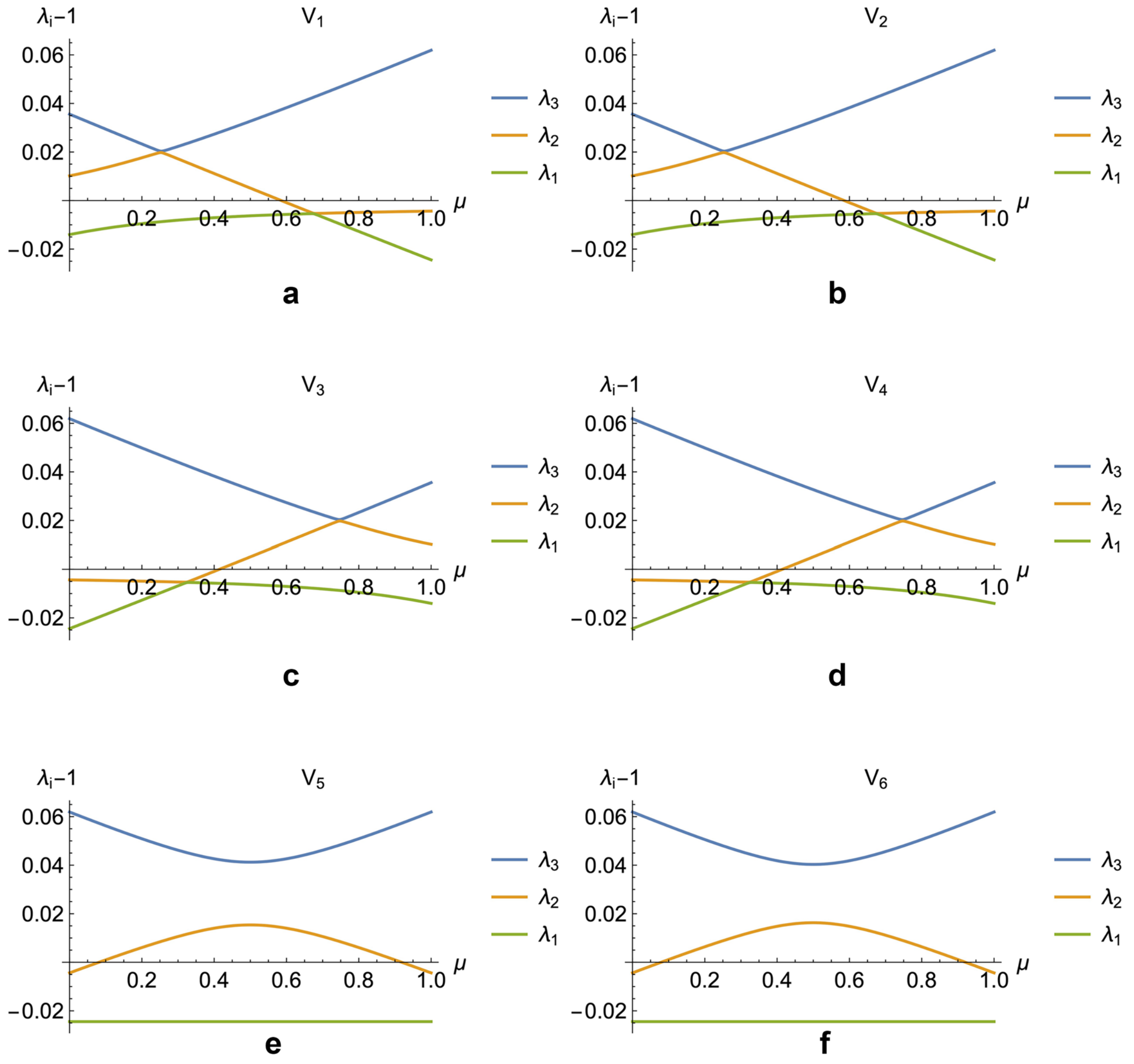

**Fig. 6** Eigenvalues $\lambda_i$ of $\boldsymbol{C}^{+}_{i,\mu}$ over volume fraction $\mu$ of a laminate between the tetragonal variants $\boldsymbol{U}_1, \boldsymbol{U}_2$. Let $\mu^*$ be the fraction such that $\lambda_1 \leq \lambda_2 = 1 \leq \lambda_3$, then the laminate with volume fraction $\mu^*$ is compatible with the orthorhombic variant $\boldsymbol{V}_i$

in energy-minimizing microstructures with sufficient freedom to allow full conversion of an established orthorhombic microstructure back to tetragonal.

If a laminate between two tetragonal variants is compatible to the initial orthorhombic phase then intermediate twinning is energetically favorable according to the adopted theory. Choosing the tetragonal variants $\mathbf{U}_1$, $\mathbf{U}_2$ (Table 1), solutions of the twinning equations

$$\boldsymbol{Q}\boldsymbol{A}^{\pm}_{\mu} - \boldsymbol{V}_i = \boldsymbol{b}_i \otimes \boldsymbol{m}_i,\ i = 1, \dots, 6 \text{ for some } \mu \in [0, 1], \quad (18)$$

where $$\boldsymbol{A}^{\pm}_{\mu} = \mu \boldsymbol{R}^{\pm}\boldsymbol{U}_2 + (1-\mu)\boldsymbol{U}_1 = \mu \boldsymbol{a}^{\pm} \otimes \boldsymbol{n}^{\pm} + \boldsymbol{U}_1, \quad (19)$$

are examined [$\boldsymbol{a}^{\pm}, \boldsymbol{n}^{\pm}$ are provided by Eq. (10)]. Solutions of (18) exist when the middle eigenvalue of $\boldsymbol{C}^{\pm}_{i,\mu} = \left(\boldsymbol{F}^{\pm}_{i,\mu}\right)^{\mathrm{T}} \boldsymbol{F}^{\pm}_{i,\mu}$ is 1, where $\boldsymbol{F}^{\pm}_{i,\mu} = \boldsymbol{A}^{\pm}_{\mu}\boldsymbol{V}^{-1}_{i}$. Plotting the eigenvalues of $\boldsymbol{C}^{+}_{i,\mu}$ ($\boldsymbol{C}^{-}_{i,\mu}$) in Fig. 6 (S2), it is shown that there exists at least one value of $\mu$ such that the laminate between $\boldsymbol{U}_1$ and $\boldsymbol{U}_2$ is compatible with every orthorhombic variant. At the transition temperature $\theta_{\mathrm{OT}} = 202^{\mathrm{o}}$, the set of minimizers is $\mathcal{M}_{\theta_{\mathrm{OT}}} = \left(\cup_{k=1}^{6} SO(3)\{\boldsymbol{V}_k, \boldsymbol{0}\}\right) \cup \left(\cup_{i=1}^{3} SO(3)\{\boldsymbol{U}_i, \boldsymbol{0}\}\right)$, which means for every orthorhombic variant there exist sequences of deformations $\mathbf{y}_k$ involving tetragonal variants such that the total potential energy is minimized as $k \to +\infty$. Therefore, the formation of intermediate twinning can be interpreted as an energetically preferred state.

In our framework, the tetragonal phase is considered an energy-minimizing state rather than an intermediate state that facilitates the transition from the cubic to the orthorhombic phase. By applying [22, Theorem 7], we find that numerous laminates between orthorhombic variants are compatible with the cubic phase, such as the laminates formed by the pairs $\boldsymbol{V}_1 - \boldsymbol{V}_2, \boldsymbol{V}_3 - \boldsymbol{V}_4, \boldsymbol{V}_5 - \boldsymbol{V}_6$. Consequently, a path exists that lowers the energy barriers without necessitating the intermediate tetragonal phase. This indicates that the tetragonal phase is indeed a minimizing state, as a direct cubic-to-orthorhombic transformation would otherwise be energetically preferable. Furthermore, XRD measurements in [18, Fig. 1] show that the tetragonal phase persists from 220 to 390 °C, providing additional evidence for the validity of the assumptions in our model.

## Incorporating Polarization in Energy Minimization

According to [19], the directions for spontaneous polarization in orthorhombic KNN configuration are the $\langle 110 \rangle_c$ directions. This is consistent with many ferroelectric or ferromagnetic martensites, in that the easy axes are eigenvectors of the transformation stretch tensors [21, 27, 35]. These equivalent crystallographic directions correspond to the eigenvectors with eigenvalues $\alpha_o$ and $\beta_o$ for each orthorhombic variant $\boldsymbol{V}_i$ (Table 2). For $\mathrm{KNN_{ex}}$, $\alpha_o \simeq 0.9899$ and $\beta_o \simeq 1.002$. Here, we choose $\beta_o$, otherwise the following structures are not predicted for the eigenvalue $\alpha_o$. From the aforementioned observations it is natural to assume that $\boldsymbol{p}_5 = p_s \widehat{\boldsymbol{p}}_5$, where $\widehat{\boldsymbol{p}}_5 = \frac{-\boldsymbol{e}_1 + \boldsymbol{e}_2}{\sqrt{2}}$, for the energy well $SO(3)\{\boldsymbol{V}_5, \pm \boldsymbol{p}_5\}$. Every orthorhombic variant $\boldsymbol{V}_i$ can be obtained through the relation $\boldsymbol{V}_i = \boldsymbol{R}_i \boldsymbol{V}_5 \boldsymbol{R}_i^{\mathrm{T}}$ and $\boldsymbol{R}_i \in \mathcal{P}(\boldsymbol{V}_5 \boldsymbol{e}_i)$. Together with $\mathbf{V}_5 \mathbf{p}_5 = \beta_o \mathbf{p}_5$ it is deduced that $\mathbf{p}_i = \mathbf{R}_i \mathbf{p}_5$ and $\mathbf{V}_i \mathbf{p}_i = \beta_0 \mathbf{p}_i$. Consequently, our assumption that polarization occurs along the direction with stretch $\beta_0$ is consistent for every variant. For convenience, the unit polarization vectors for the variants $\boldsymbol{V}_5, \boldsymbol{V}_6, \boldsymbol{V}_3$ and $\boldsymbol{V}_4$ involved in the crossing twin of Fig. 4, are given explicitly by

$$\widehat{\boldsymbol{p}}_5 = \frac{-\boldsymbol{e}_1 + \boldsymbol{e}_2}{\sqrt{2}}, \widehat{\boldsymbol{p}}_6 = \frac{\boldsymbol{e}_1 + \boldsymbol{e}_2}{\sqrt{2}}, \widehat{\boldsymbol{p}}_3 = \frac{-\boldsymbol{e}_1 + \boldsymbol{e}_3}{\sqrt{2}}, \widehat{\boldsymbol{p}}_4 = \frac{\boldsymbol{e}_1 + \boldsymbol{e}_3}{\sqrt{2}}. \tag{20}$$

Here, it is implied $\mathbf{p}_i = p_s \widehat{\mathbf{p}}_i$, where $p_s$ accounts for saturation. In the orthorhombic phase, $\theta_o = 187^o$, energy wells are described by the set $\mathcal{M}_{\theta_o} = \cup_{i=1}^{6} SO(3)\{\boldsymbol{V}_i, \pm \boldsymbol{p}_i\}$. Now let $\{\nabla \mathbf{y}(\mathbf{x}), \mathbf{p}(\mathbf{y})\} \in \mathcal{M}_{\theta_o}$ for all $\boldsymbol{x} \in \Omega, \Omega$ is the reference configuration (cubic state). Then under the formation of the crossing twin, the only contribution to the total energy of the electrostatic configuration is due to the depolarization energy, specifically the total energy is

$$E[\boldsymbol{y}, \boldsymbol{p}, \theta] = \frac{1}{2} \int_{R^3} |\nabla \phi(z)|^2 \mathrm{d}z. \tag{21}$$

The depolarization energy vanishes when $\mathrm{div}\,\boldsymbol{p} = 0$, and it is divergence free if the polarization jump conditions (8) are satisfied at the interfaces between the variants. Surprisingly, choosing signs with order $\pm\mathbf{p}_5, \mp\mathbf{p}_6, \pm\mathbf{p}_3, \mp\mathbf{p}_4$, condition (8) holds. Hence, depolarization energy is minimized in the interior of the body. To validate this observation one needs to notice first that the specific deformation of (17) implies

$$\boldsymbol{n}_1 = -\boldsymbol{n}_3 = \boldsymbol{e}_1,\ \boldsymbol{n}_4 \cdot \boldsymbol{e}_3 = \boldsymbol{n}_2 \cdot \boldsymbol{e}_2,\ \boldsymbol{n}_4 \cdot \boldsymbol{e}_2 = \boldsymbol{n}_2 \cdot \boldsymbol{e}_3,\ \boldsymbol{n}_2 \cdot \boldsymbol{e}_3 = -\boldsymbol{n}_2 \cdot \boldsymbol{e}_2, \tag{22}$$

where conditions (15) hold for every $\boldsymbol{n}_i$. Then for every interface between two orthorhombic variants condition (8) is satisfied, namely

$$(\boldsymbol{p}_5 + \boldsymbol{p}_6) \cdot \boldsymbol{n}_1 = (\boldsymbol{p}_6 + \boldsymbol{p}_3) \cdot \boldsymbol{n}_2 = (\boldsymbol{p}_3 + \boldsymbol{p}_4) \cdot \boldsymbol{n}_3 = (\boldsymbol{p}_4 + \boldsymbol{p}_5) \cdot \boldsymbol{n}_4 = 0. \tag{23}$$

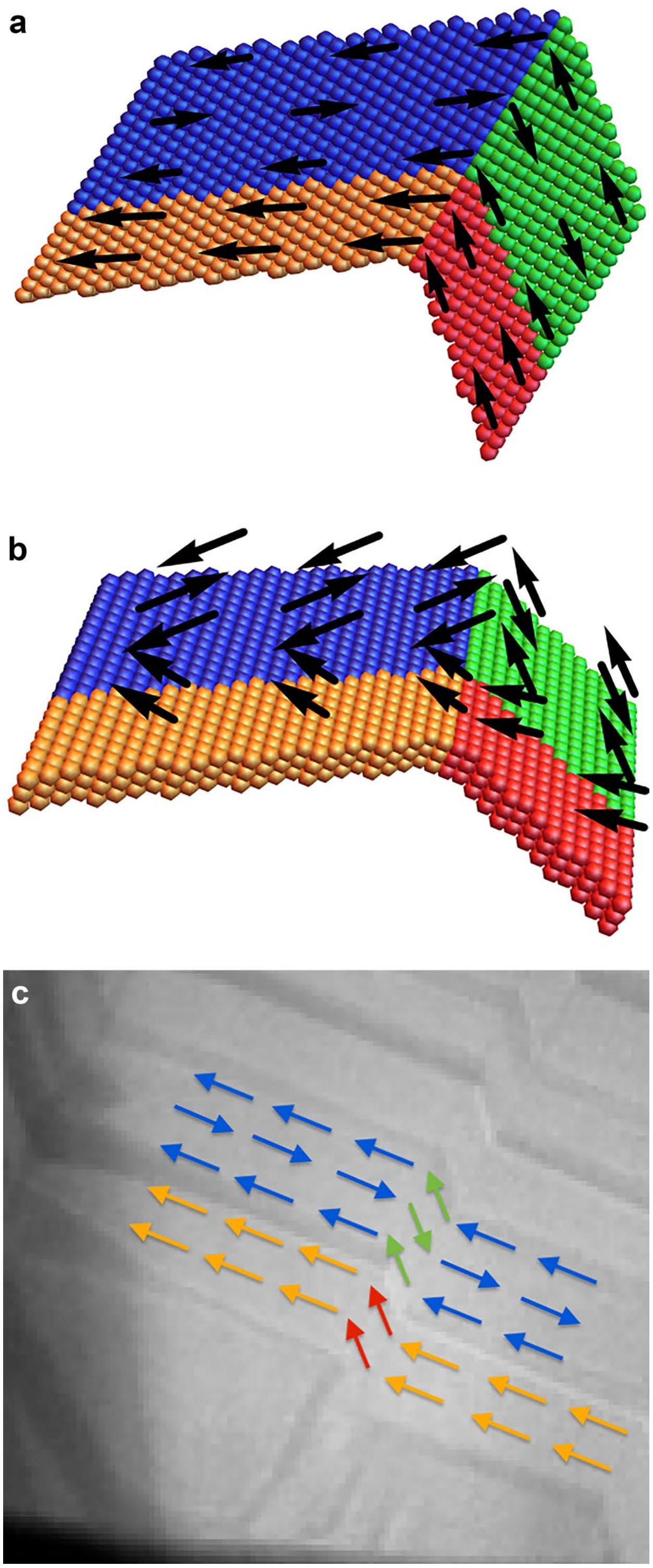

**Fig. 7** **a** and **b** Different angles of spontaneous polarization in top of variants $\boldsymbol{V}_5$, $\boldsymbol{V}_6$, $\boldsymbol{V}_3$and$\boldsymbol{V}_4$ (green, blue, orange, and red). Here, $180^o$ domains are formed within $\boldsymbol{V}_5$ and $\boldsymbol{V}_6$ variants, proposing an underlying mechanics for the formations of the observed narrow bands of Fig. 5a. We have chosen the compatible polarization vectors $\boldsymbol{p}_5 = p_s \widehat{\boldsymbol{p}}_5, \boldsymbol{p}_6 = -p_s \widehat{\boldsymbol{p}}_6, \boldsymbol{p}_3 = p_s \widehat{\boldsymbol{p}}_3$ and $\boldsymbol{p}_4 = -p_s \widehat{\boldsymbol{p}}_4$, $\widehat{\boldsymbol{p}}_i$ is given from Eq. (20). **c** The corresponding polarization vectors drawn on the observed orthorhombic phase. Green, blue, orange, and red arrows show the polarization vectors in $\boldsymbol{V}_5$, $\boldsymbol{V}_6$, $\boldsymbol{V}_3$and$\boldsymbol{V}_4$ phases, respectively (Color figure online)

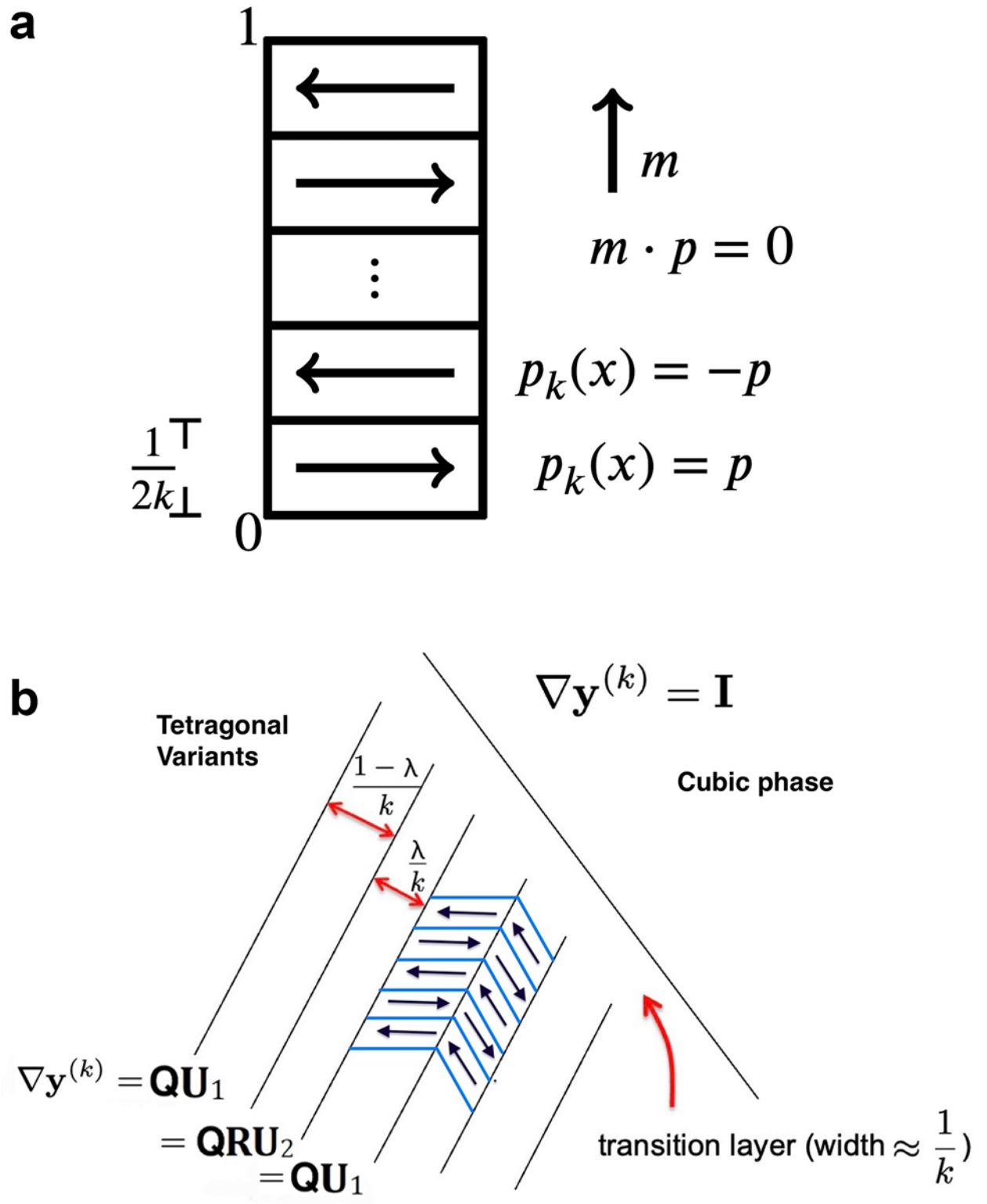

**Fig. 8** **a** A sequence $\boldsymbol{p}_k$ forming 180° domains along $\mathbf{m}$, where $\mathbf{m}$ is the normal to the polarization vector $\mathbf{p}$. The region is subdivided into $2k$ subregions representing domains. **b** A laminate between the tetragonal variants $\boldsymbol{U}_1$ and $\boldsymbol{U}_2$ compatible with the cubic phase. Within each tetragonal variant polarization vectors as in **a** are formed, parallel blue lines, with the additional property that are divergence free across the twin interface between $\boldsymbol{QRU}_2$ and $\boldsymbol{QU}_1$ (Color figure online)

Note that the above equations do not imply that depolarization energy is zero due to possible contributions of $\mathrm{div}\boldsymbol{p}_i$ at the boundary of the grain.

Now let the narrow bands in Fig. 5a represent 180° domains within a variant, for example, the crossing twin microstructure with their respective polarization vectors are illustrated in Fig. 7. But why are these 180° domains energetically preferable? First, one can notice that $\mathrm{div}\boldsymbol{p} = 0$ still holds in the interior of each variant, i.e., there is no contribution to the depolarization energy from the bulk. Furthermore, the formation of these domains can be considered as a part of a sequence minimizing the total energy. For simplicity, assume the homogeneous deformations $\boldsymbol{y}(\boldsymbol{x}) = \boldsymbol{F}\boldsymbol{x}$ is performed on $\Omega$ and let $\{\mathbf{F}, \pm\mathbf{p}\} \in \mathcal{M}_\theta$. Let $\mathbf{m}$ be normal to $\mathbf{p}$ with $|\mathbf{m}| = 1$ and $\boldsymbol{p}_k$ the sequence depicted in Fig. 8a. It has been shown [21, Proposition 5.1], that as $k \to \infty$ the depolarization energy goes to zero. This can be considered as a macroscopic zero polarization that cancels energy contributions from the boundary. This is a possible mechanics for domains formation of this kind and suggests that different geometries should affect domains structure.

In the tetragonal phase, the elongation of the unit cell conditions preferred $\langle 100 \rangle_c$ directions for the spontaneous polarization direction [20]. Then, each tetragonal variant is related by the polarization direction through

$$\boldsymbol{U}_i = \alpha_t 1 + (\gamma_t - \alpha_t)\hat{\boldsymbol{p}}_i \otimes \hat{\boldsymbol{p}}_i, \quad \text{where } \hat{\boldsymbol{p}}_i = \boldsymbol{e}_i \quad \text{and } \boldsymbol{p}_i = p_s\hat{\boldsymbol{p}}_i. \tag{24}$$

For the measured lattice parameters (Table S1), two distinct tetragonal variants are rank-one connected. In pure tetragonal phase, the energy wells are described by $\mathcal{M}_{\theta_T} = \cup_{i=1}^{3} SO(3)\{\boldsymbol{U}_i, \pm\boldsymbol{p}_i\}$. As before, if $\boldsymbol{U}_j = \boldsymbol{R}\boldsymbol{U}_k\boldsymbol{R}^{\mathrm{T}}$ then $\boldsymbol{p}_j = \boldsymbol{R}\boldsymbol{p}_k, \boldsymbol{R} \in \mathcal{P}(\boldsymbol{U}_k\boldsymbol{e}_i)$ and condition (8) is satisfied. For instance, take the twin formed from $\boldsymbol{U}_1$ and $\boldsymbol{U}_2$. The two possible interfaces have normals $\boldsymbol{n}^{\pm} = (1, \pm 1, 0)$, resulting to $(\boldsymbol{p}_2 - \boldsymbol{p}_1) \cdot \boldsymbol{n}^+ = 0$ and $(\boldsymbol{p}_2 + \boldsymbol{p}_1) \cdot \boldsymbol{n}^- = 0$, which implies there is no depolarization energy in the interior of the twin. Then, following the proof of [21, Theorem 5.2], there exists a minimizing sequence $\{\boldsymbol{y}_k, \boldsymbol{p}_k\}$ such that the total electro-elastic energy is minimized as $k \to +\infty$. A sequence of this form can be observed as well under the cubic to tetragonal phase transition. An example of an arbitrary cubic to tetragonal transitions is illustrated in Fig. 8b. In that region a sequence of 180° domains within each tetragonal variant emerge, with pole-free interfaces. On the other hand, in the transition layer depolarization energy is zero if the polarization vectors within this layer are parallel to the phase boundary (boundary of cubic phase) and additionally divergence free with the polarization vectors arising inside the tetragonal variants. Intuitively, it is very unlikely to produce a pole-free transition layer. However, it has been shown in [21] that the depolarization in the transition layer vanish when $k \to +\infty$, where oscillation for both strains and polarization vectors occur. Here, the polarization vectors do not need to oscillate much finer compared to the strain oscillations. Another example of sequences with these properties can be seen in [21, Fig. 1].

In electrostriction, elastic deformations affect spontaneous polarization directions and vice versa, i.e., variants

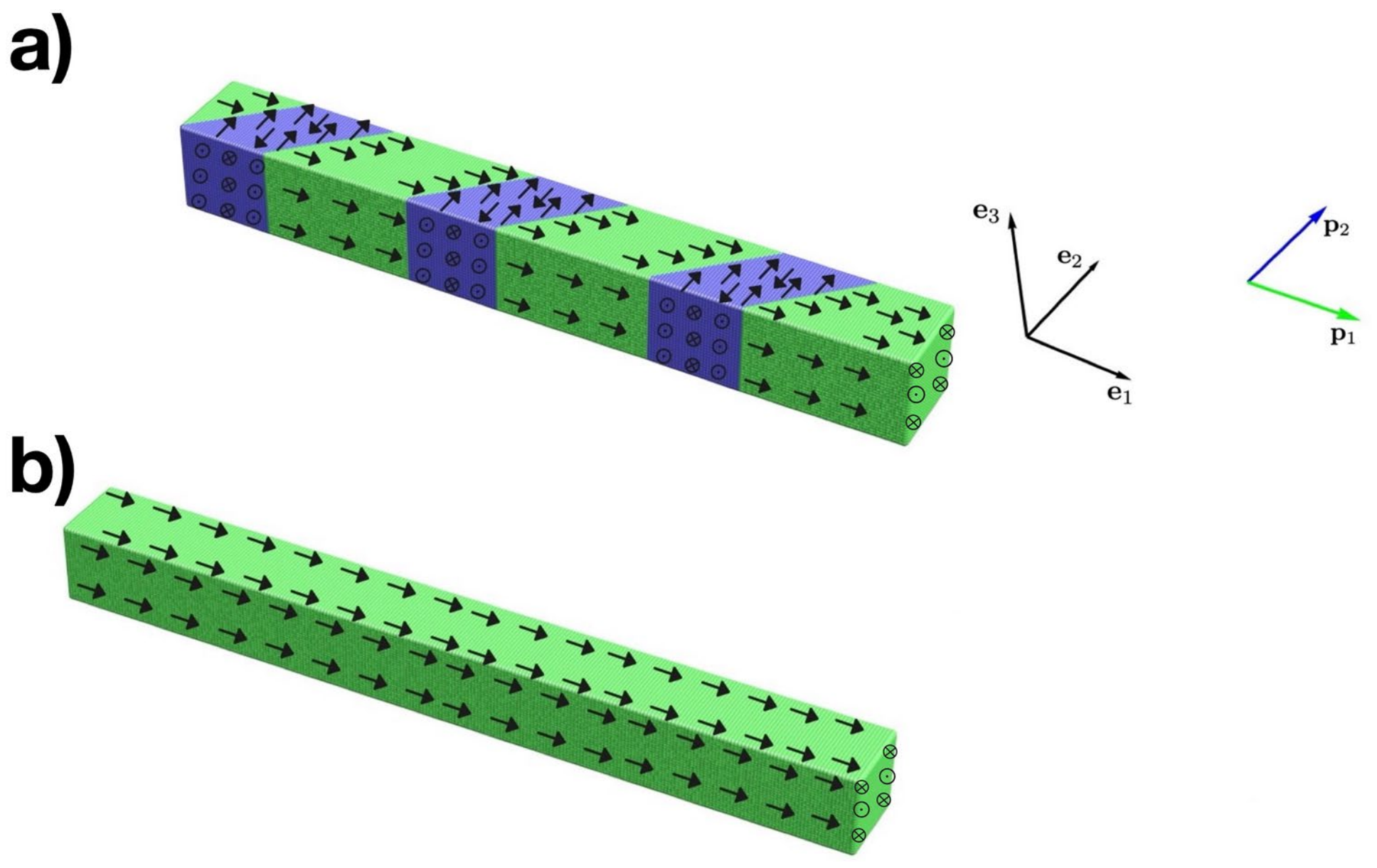

**Fig. 9** Depolarization energy affects microstructures, an example in cubic to tetragonal transformation of a long slab over the $\boldsymbol{e}_1$ axis. The symbols ⊗, ⊙ indicate that a vector is going into, coming out from a plane, respectively. **a** When two compatible tetragonal variants emerge, $\boldsymbol{RU}_2$ (blue regions) and $\boldsymbol{U}_1$(green regions) with their corresponding polarization vectors (black arrows) along directions $\boldsymbol{p}_2 = \boldsymbol{Re}_2$ and $\boldsymbol{p}_1 = \boldsymbol{e}_1$, the elastic deformation is fully compatible, see section "The Cubic-to-Tetragonal Transformation," but the depolarization energy is non-zero. Poles are formed on the blues surfaces which are normal to $\boldsymbol{e}_2$ and on the green surfaces that are normal to $\boldsymbol{e}_1$. **b** A state with lower total energy when only variant $\boldsymbol{U}_1$ appear. Then, energy is non-zero at the plane perpendicular to $\boldsymbol{e}_1$ where poles are formed. In the latter case simpler 180$^o$ domains create minimizing sequences of the depolarization energy, while in **a** (Color figure online)

selection is influenced by preferred polarization directions. An example highlighting this feature is shown in Fig. 9. A configuration with a single tetragonal variant of $KNN_{ex}$ is preferable than a compatible twin, generating fewer poles which means lower depolarization energy. Note that there are no energy contributions from elastic incompatibilities in either case. This shows that the energy barriers, under phase transformations, are altered from the geometry of the material. A more systematic study of the aforementioned observations is left as future work, where one could examine polarization compatibility and how energy barriers evolve in the electrostrictive setting, analogously to the pure elastic case [53].

Therefore, the experimentally observed directions of spontaneous polarization in both the orthorhombic and tetragonal phases [19, 20] and their phase transitions can be interpreted as energy-minimizing directions through the adopted nonlinear setting. In the absence of any other comparable experimental data, we avoid to present more details about these sequences.

## Conclusion

Within this work, we employ a full nonlinear electroelastic energy for modeling the recently observed intermediate twinning, first-order phase transitions and spontaneous polarization in $KNN_{ex}$. The nonlinear theory has been developed in the framework of electrostriction and geometrically nonlinear elasticity. We show that microstructures observed in the cubic to tetragonal and tetragonal to orthorhombic phase transitions are energetically favorable through minimizing sequences $\{\boldsymbol{y}_k\}$, where the total free energy is minimized as $k \to +\infty$. These sequences correspond to simple or complex laminates among variants of the involved phase and $k \to +\infty$ represents fine phase mixtures. Intermediate twinning is interpreted as a laminate of this type, but departing from common practice, the laminate contains the higher symmetry tetragonal variants and it is compatible to a lower symmetry orthorhombic variant. The most striking agreement between theory and experiments occurs in the pure orthorhombic phase where crossing twins arise. Four interfaces separating four distinct orthorhombic variants intersect along a line. The agreement between theoretical predicted and experimentally observed angles is remarkable. With respect to the spontaneous polarization assuming that the polarization directions coincide with a stretch directions of the correlated phase the depolarization energy is minimized. Despite complex stoichiometries and transition paths in $KNN_{ex}$, the phase transition and domain dynamics obey energy minimization at all times and can thus be manipulated along the set of presented criteria. We believe that the nonlinear electroelastic theory can serve as a powerful tool in understanding, exploring and tailoring the electromechanical properties of complex ferroelectric ceramics.

**Funding** Open access publishing provided by King Abdullah University of Science and Technology (KAUST). The work of GG, P-L P-G, and RDJ was partly supported by a Vannevar Bush Faculty Fellowship (Grant No. N00014-19-1-2623). The initial work of P-L P-G and EQ was funded by the DFG via the Reinhart Koselleck Project “Crystallographically compatible ceramic shape memory materials” (Project Number 313454214).

**Data Availability** The data that support the findings of this study are available from the corresponding author upon reasonable request.

**Declarations**

**Competing Interests** The authors declare that they have no known competing financial interests or personal relationships that could have appeared to influence the work reported in this paper.

**Publisher's Note** Springer Nature remains neutral with regard to jurisdictional claims in published maps and institutional affiliations.